\documentclass[10pt]{article} % For LaTeX2e
\usepackage[preprint]{tmlr}

\usepackage{amsmath,amsfonts,bm}

\def\eqref#1{equation~\ref{#1}}
\def\1{\bm{1}}

\DeclareMathAlphabet{\mathsfit}{\encodingdefault}{\sfdefault}{m}{sl}
\SetMathAlphabet{\mathsfit}{bold}{\encodingdefault}{\sfdefault}{bx}{n}

\usepackage{hyperref}
\usepackage{url}
\usepackage[pdftex]{graphicx}

\title{SYMBIOSIS: Systems Thinking and Machine Intelligence for Better Outcomes in Society}

\author{\name Sameer Sethi \email sethis@google.com \\
      \addr Google Research
      \AND
      \name Donald Martin, Jr. \email dxm@google.com \\
      \addr Google Research
      \AND
      \name Emmanuel Klu \email eklu@google.com\\
      \addr Google Research
      }

\def\month{MM}  % Insert correct month for camera-ready version
\def\year{YYYY} % Insert correct year for camera-ready version
\def\openreview{\url{https://openreview.net/forum?id=XXXX}} % Insert correct link to OpenReview for camera-ready version

\begin{document}

\maketitle

\begin{abstract}
This paper presents \textit{SYMBIOSIS}, an AI-powered framework and platform designed to
make Systems Thinking accessible for addressing societal challenges and unlock paths for leveraging systems thinking frameworks to improve AI systems. The platform establishes a centralized, open-source repository of systems thinking/system dynamics models categorized by Sustainable Development Goals (SDGs) and societal topics using topic modeling and
classification techniques. Systems Thinking resources, though critical for articulating causal theories in complex problem spaces, are often locked behind specialized tools and intricate notations, creating high barriers to entry. To address this, we developed a generative co-pilot that translates complex systems representations - such as causal loop and stock-flow diagrams - into natural language (and vice-versa), allowing users to explore and build models without extensive technical training. 

Rooted in community-based system dynamics (CBSD) and informed by community-driven insights on societal context, we aim to bridge the problem understanding chasm. This gap, driven by epistemic uncertainty, often limits ML developers who lack the community-specific knowledge essential for problem understanding and formulation, often leading to ill informed causal assumptions, reduced intervention effectiveness and harmful biases. Recent research identifies causal and abductive reasoning as crucial frontiers for AI, and Systems Thinking provides a naturally compatible framework for both. By making Systems Thinking frameworks more accessible and user-friendly, SYMBIOSIS aims to serve as a foundational step to unlock future research into responsible and society-centered AI that better integrates societal context by leveraging systems thinking frameworks and causal modeling methods. Our work underscores the need for ongoing research into AI's capacity to understand essential characteristics of complex adaptive systems - such as feedback processes and time delays - paving the way for more socially attuned, effective AI systems. 

\end{abstract}

\section{Introduction}

As artificial intelligence (AI) continue to transform domains ranging from healthcare to education to social services, there is an increasing recognition of the limitations that come with operating these systems in isolation from societal and contextual understanding. Systems dynamics (SD) and systems thinking (ST) methodologies have long provided a framework for understanding complex, interconnected problems. However, despite their capacity to capture inter-dependencies, feedback loops, and non-linearities that characterize real-world challenges, these tools have remained largely inaccessible to many stakeholders due to specialized notations, complex model representations, and the proprietary nature of certain tools and models. This paper presents \textit{SYMBIOSIS}, an ML-driven framework and platform designed to bridge these gaps by making systems thinking models accessible and interpretable, using a centralized repository and AI-based tools to support diverse communities in creating, exploring, and using these models.

Recent research has identified causal and abductive reasoning as central to advancing AI's effectiveness in addressing complex, socio-technical problems (\cite{larson2021myth}). Systems Thinking, with its inherent emphasis on causal structures, feedback, and iterative problem-solving, aligns well with these reasoning methodologies, making it an ideal candidate for integration with AI to support interpretive, context-aware solutions in high-stake domains like healthcare, public policy, and environmental sustainability (\cite{barton2006fresh}).

\section{Background and Motivation}

AI systems must ultimately safely operate in the broader societal context in which they will be deployed (\cite{selbst2019fairness}). Societal context can be defined as the dynamic and complex collection of social, cultural, historical, political, economic and environmental circumstances. Societal context is another way to conceptualize society as a whole and can be categorized as a complex adaptive system (\cite{martin2020extending}).  Complex adaptive systems (CAS) are complex, non-linear, adaptive and comprised of feedback loops and time delays. Societal contexts generate both the problems that motivate the development of AI systems and the data that is essential for training them.  They are the data generating process that fuels statistical methods. Due the dynamically complex nature of societal context developers of AI systems and AI-based products tend to abstract away and ignore it. These abstraction practices result in epistemic uncertainty or lack of knowledge of the societal context on behalf of AI developers and the systems they produce. This epistemic uncertainty is a key root cause of harmful societal bias as well as fragility in AI systems. Additionally, it undermines the ability to leverage AI to solve complex societal problems.  System dynamics is a mature and promising method for helping humans and AI systems understand and reason about societal contexts and the dynamically complex problems they generate. 

\subsection{Role of Systems Thinking and System Dynamics in Complex Problem Solving}

Systems Thinking, encompassing methodologies such as system dynamics, offers a structured approach to understanding and modeling complex systems. Unlike traditional analysis, which often focuses on linear causation, Systems Thinking emphasizes feedback loops, delays, and accumulations that reveal the broader dynamics underlying systemic issues. This is critical in domains such as healthcare, criminal Justice, and social policy, where factors influencing outcomes are deeply interconnected via feedback loops.  A classic example of how feedback loops can exacerbate disparities can be found in predictive policing in which predictions of crime locations based on historical data lead to over policing of said locations resulting in skewed arrest rates and amplified predictions of crimes that do not match the true crime rate (\cite{ensign2018runaway}). 

Despite its utility, Systems Thinking has been limited by accessibility challenges. Models in System dynamics, while valuable, are often stored in isolated repositories, and the tools required to access and use these models are proprietary, requiring specialized knowledge that many community practitioners lack. \textit{SYMBIOSIS} addresses this limitation by providing a centralized, open-access repository of systems models, along with a generative AI-powered co-pilot that translates complex model representations into natural language, enabling users from diverse backgrounds to explore and build upon systems-thinking models without deep technical training. This work aligns with ongoing efforts to create transparent, interpretable models that include the lived experiences and insights of community stakeholders, which have been shown to enhance AI's fairness and effectiveness in real-world applications (\cite{Martin_2023}).

\subsection{Bridging the Epistemic Uncertainty and Problem Understanding Gap in ML}

In high-stakes application of ML, epistemic uncertainty arises when developers lack sufficient understanding of the socio-technical contexts in which their models will be applied. This knowledge gap can lead to "problem understanding chasms", where the views of AI practitioners, who often approach problems from a data-centric perspective, diverge from those of community members who possess rich, qualitative knowledge about the local context of the issues (\cite{Martin_2023}). For instance, in healthcare, racial biases in predictive models have often stemmed from a lack of understanding of the socio-historical factors influencing health disparities, resulting in models that fail to accurately address the needs of the marginalized groups (\cite{obermeyer2019dissecting}).

\textit{SYMBIOSIS} aims to bridge this problem understanding gap by making system thinking models more accessible, thereby enabling stakeholders with deep contextual knowledge to participate in model exploration and generation. The platform is rooted in community-based system dynamics (CBSD), a participatory approach that engages communities in co-creating causal theories and system insights. CBSD has been used effectively in domains like public health and social services, where understanding local context is essential for building effective, fair solutions (\cite{hovmand2014group}). By leveraging AI to make Systems Thinking tools interpretable and engaging for non-experts, \textit{SYMBIOSIS} prompts a symbiotic relationship between ML developers and community stakeholders, supporting a more inclusive problem formulation phase that can mitigate epistemic uncertainty.

\subsection{Systems Thinking as a Framework for Causal and Abductive Reasoning in AI}

Recent advancements in AI research underscore the need for models that can reason causally and abductively (\cite{larson2021myth}) to tackle complex problems with incomplete information. Causal reasoning allows AI to consider not just correlations but also potential cause-and-effect relationships, while abductive reasoning supports the generation of plausible explanations based on observed data (\cite{walton2014abductive},\cite{paul1993approaches}) . Systems Thinking inherently supports these forms for reasoning through its focus on causal structure and iterative hypothesis testing, making it particularly valuable for AI systems aiming to operate in dynamic and uncertain environments.

Systems models, such as causal loop diagrams (CLDs) and stock-and-flow diagrams (SFDs), capture the interconnected elements and feedback mechanisms in complex systems, facilitating both causal inference and abductive reasoning about system behaviours over time. When integrated with AI, these systems-thinking tools could enable models to consider not only the immediate effects of interventions but also the long-term impacts, feedback, and delayed effects. However, there are limitations to how well AI can currently interpret and implement causal structures, and these limitations highlight the need for further research and development in this area.

\subsection{AI for Systems Thinking as the foundation for future research on Systems Thinking for AI}

The preceding sections illustrate the long-term vision and our motivation for \textit{SYMBIOSIS}, where the integration of Systems Thinking with AI methods can fundamentally enhance the problem understanding and formulation, causal and abductive reasoning and interpretability of AI systems. However, this paper represents the initial phase of this journey: leveraging AI to make Systems Thinking accessible. The primary challenge lies in the barrier represented by Systems Thinking/System Dynamics notations and specialized models, which often require extensive training and deep familiarity with SD principles. In our experience, this complexity demands significant effort to align AI practitioners with the frameworks and language of Systems Thinking, delaying exploration into its potential to enrich AI methods. Our work in \textit{SYMBIOSIS} aims to address this challenge by creating an accessible entry point for ML practitioners, allowing them to interact with systems-thinking models and explore dynamic interdependencies in societal issues without needing prior expertise in SD notation or formal systems-thinking frameworks.

This foundational approach is essential to establish the critical first stage in AI development workflow: problem understanding and formulation. Problem formulation is a critical stage, especially for AI systems with complex socio-technical applications. Yet, current AI workflows often lack structured methods to incorporate contextual knowledge and community perspectives that embedded in existing systems-thinking models. With \textit{SYMBIOSIS}, ML practitioners gain a resource that allows them to explore existing community knowledge on complex issues, articulated through causal relationships and feedback mechanisms, and presented in an accessible format. This removes the need for practitioners to first gain fluency in specialized SD notations and models before engaging with problem context, enabling them to begin directly with the problem articulation, exploration, and hypothesis stages.

Furthermore, \textit{SYMBIOSIS} provides an interactive environment where practitioners can engage in problem formulation - defining the "what", "why", and "how" of the systems they aim to develop in natural language. Through the platform's generative co-pilot, these inputs can be translated into systems-thinking models that encapsulate the causal structure, delays, and feedback loops inherent to complex problems. This step-by-step process supports the creation of structured systems models that go beyond conventional problem statements, enabling practitioners to define interventions, simulate potential outcomes, and iterate based on model feedback. By removing the barriers posed by systems dynamics language and notations, \textit{SYMBIOSIS} empowers ML practitioners to use the platform as a bridge to articular problem formulations that are both context-rich and systems-oriented.

\section{System Architecture}

\begin{figure}[ht]
    \centering
    \includegraphics[width=\textwidth]{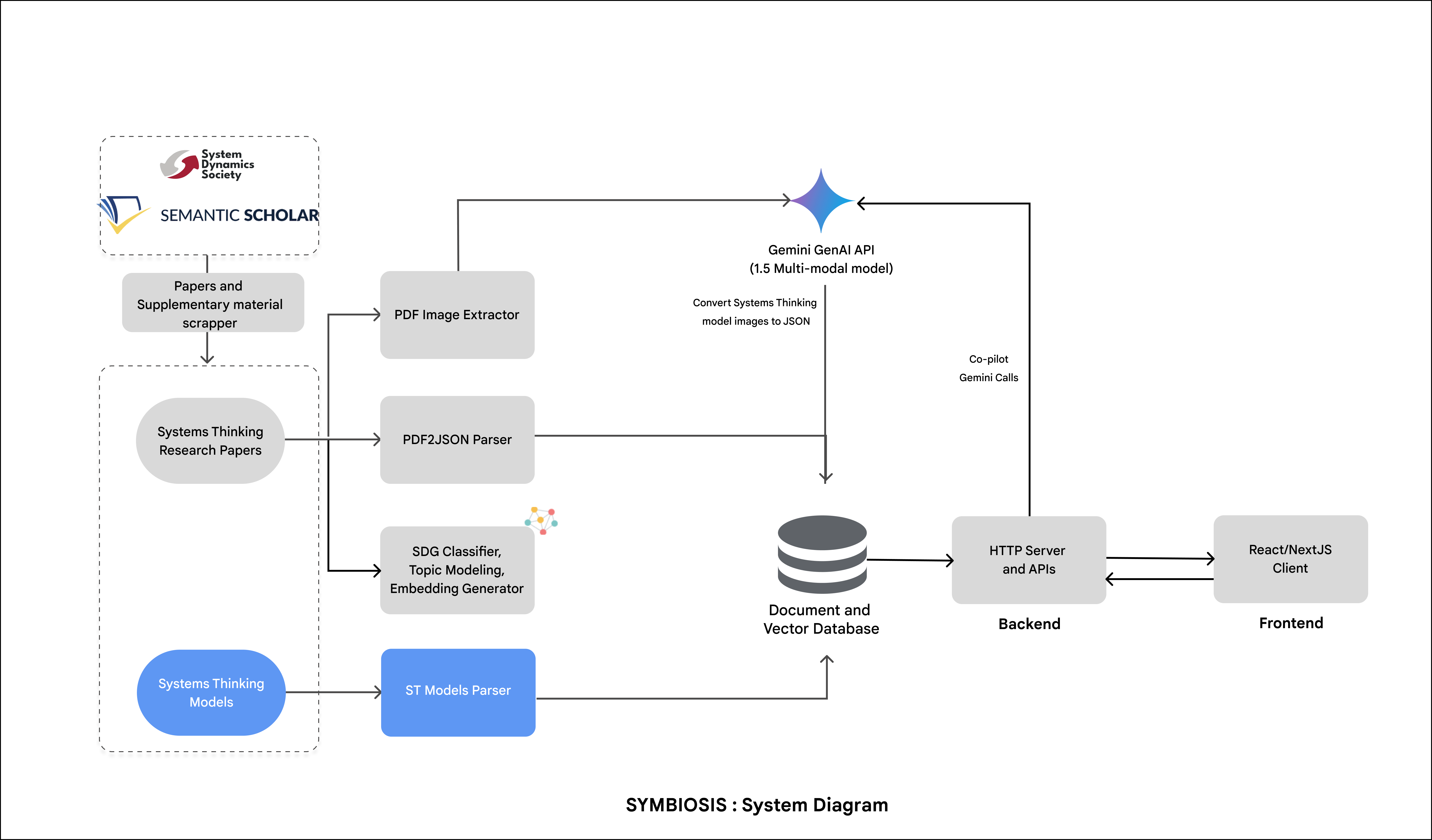}
    \caption{High-level System Architecture}
    \label{fig:architecture}
\end{figure}

The \textit{SYMBIOSIS} platform is built to bridge Systems Thinking and AI through an integrated, user-accessible interface that supports exploration, interaction, and model development for complex societal challenges. At a high-level the platform operates with a three-layer architecture (Figure \ref{fig:architecture}) designed for scalability, efficient data processing, and user accessibility. The architecture consists of:

\begin{enumerate}
    \item \textbf{A Repository Layer} based on an Elasticsearch database (\cite{gormley2015elasticsearch}) for fast and flexible storage and retrieval of index research papers, transformed systems thinking models and embedding vectors.
    \item \textbf{An HTTP-based API Layer} implemented with FastAPI (\cite{lubanovic2023fastapi}), which serves as the backend for managing data requests, processing queries, and interfacing between the repository and frontend layers.
    \item \textbf{A Frontend Layer} developed with Next.js (\cite{thakkar2020next}), which provides an interactive user interface and includes middleware to handle pre-rendering, caching and session management.
\end{enumerate}

The architecture supports a seamless interaction workflow, allowing users to access, explore, and interact with complex systems thinking models via an intuitive frontend interface that draws on backend processing and a robust repository infrastructure.

\subsection{Repository of research papers and models}

At the core of \textit{SYMBIOSIS} is a repository layer built on Elasticsearch, a powerful search and analytics engine known for its scalability, full-text search capabilities and efficient indexing. Elasticsearch is used to store and organize both systems thinking models and research papers, providing a highly flexible structure for managing these heterogeneous data types. This layer allows for rapid retrieval of content based on complex queries, which is essential for search functionality, fetching model data for rendering or fetching data to use in generative AI functions.

Elasticsearch's indexing capabilities are leveraged to perform structured and unstructured searches across the repository, making it possible for users to search by keywords, SDG classifications, thematic topics, and even specific systems thinking elements such as feedback loops or causal relationships. By enabling efficient querying and indexing, the repository layer reduces data latency, ensuring quick response times for users interacting with the system through complex, multi-faceted searches.

\subsection{HTTP-based Service API Layer}

The HTTP-based service API layer, developed with FastAPI as the API framework and Uvicorn as the ASGI web server, serves as the backend interface for managing data requests between the frontend and Elasticsearch repository. FastAPI is chosen for its asynchronous support and performance optimization, which enables high-throughput request handling with minimal latency. This layer is responsible for managing the logic required to process incoming user queries, retrieve data from Elasticsearch, and perform additional backend processing tasks such as topic modeling and SDG classification.

\subsection{Frontend Layer}

The frontend layer, built with Next.js, delivers an intuitive and highly interactive user interface for accessing Symbiosis's features. Next.js, a React-based framework, supports both server-side and client-side rendering, enabling dynamic content generation and high performance across various user devices. The frontend includes serveral core components, such as model exploration interfaces, search functionality, and the generative AI-powered co-pilot for natural language interactions with systems thinking models.

\subsection{Transforming Research Papers and Models to Structured Data}

A critical component of the \textit{SYMBIOSIS} platform is the ability to transform unstructured research papers and systems models into structured, searchable data. To achieve this, we developed a custom XMILE (\cite{eberlein2013xmile}) parser tailored to process and extract relevant data from the supplementary files and models associated with research papers in publicly available databases. This transformation process enables the integration of causal relationships, model parameters, and contextual metadata into a structured format that can be readily indexed and queried within the \textit{SYMBIOSIS} platform.

To facilitate the broader use of the platform and encourage reproducibility, we will be open-sourcing data pipeline code for transforming and standardizing systems thinking models and research papers. These data pipelines automate the process of converting raw XMILE files and research papers into structured data, following best practices for data validation, error handling, and schema mapping.

\subsection{SDG classification and Topic Modeling}

To organize and categorize the diverse research papers, we implemented a dual approach of SDG classification and topic modeling. This approach leverages both traditional machine learning techniques and generative AI to accurately align resources with relevant SDGs and core thematic topics.

For SDG classification, we utilized a Long Short-Term Memory (LSTM)-based multi-label classifier to identify which SDGs each paper most closely aligns with. The LSTM model, known for its sequence processing capabilities, effectively captures context within the text, enabling accurate classification across complex, multi-topic documents.  A sequential Keras model with a single LSTM layer and a dropout of 0.5 was trained for 10 epochs on labeled abstracts from the Aurora project ( \citep{kashnitsky2023evaluating}), using an AdamW optimizer at a learning rate of 0.001, a binary cross-entropy loss function, a batch size of 64, and an early stopping patience of 3 epochs. Micro-average metrics for all SDG objective and target labels reported 0.86 precision, 0.67 recall and an F1 score of 0.75. 

For topic modeling, we built a BertTopic model that leveraged a Gemini 1.5 Flash model as its representation model to improve label quality. We first generated embeddings of the full text of SD papers using the Gemini Text Embedding model. Given the large size of the embedding space, we reduced dimensionality using UMAP, then performed clustering using HDBSCAN. Lacking a standardized set of topics to label with, we applied an unsupervised approach by prompting the Gemini model to generate topic labels for the embedding clusters which were tokenized into keywords. 

\subsection{Use of Generative AI}

We prompt-tuned a Gemini 1.5 Flash model to generate structured representations (JSON) from images of causal loop diagrams (CLDs). With few shot examples, the model was instructed to identify all variables, causal links, and causal loops in an image. We provided definitions of these elements, and described visual aids that could be used to identify them. For example, we mentioned that each causal link would have an arrow with the 'from' variable at the arrow's start, and the 'to' variable at the arrow's head. 

To improve robustness, we added instructions to ignore stylistic properties like colors, and used a chain-of-thought (CoT) approach that required the model to provide its reasoning for the elements it identified. We additionally provided examples of CLD images with distinctly colored bounding boxes around each element, and asked the model to complete an object detection task as part of its reasoning. The outputs from the model included variable names, pairs of variables with directional causal links and their polarities, and all loops identified, reinforcing or balancing. 

\section{Using SYMBIOSIS}

The \textit{SYMBIOSIS} platform offers a range of interactive tools designed to make Systems Thinking accessible and actionable for users from diverse backgrounds. Through its intuitive interface, users can explore systems models, access AI-driven insights, generate new perspectives on complex societal issues. The following subsections showcase key features, including the model explorer, and the generative AI co-pilot for query-based assistance. Each feature is illustrated with screenshots to demonstrate how they enable users to navigate, interpret, and build upon systems thinking models effectively.

\subsection{Landing Page}

\begin{figure}[ht]
    \centering
    \includegraphics[width=\textwidth]{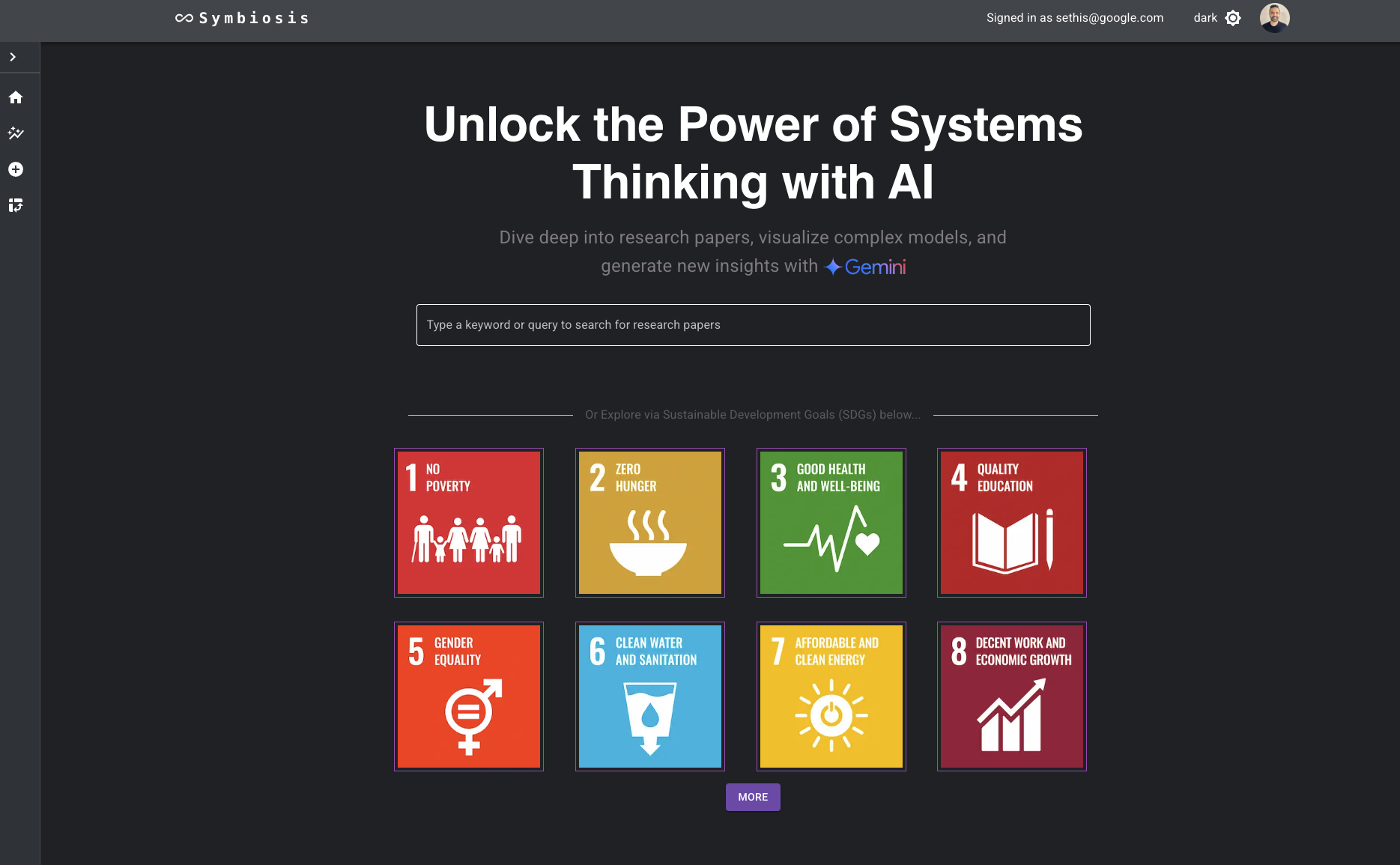}
    \caption{SYMBIOSIS Landing Page}
    \label{fig:landing_page}
\end{figure}

The landing/home of the platform (as shown in Figure \ref{fig:landing_page}) provides access to a left navigation panel to access different parts of the platform such as Explorer, Create and Table View pages. In the landing page we provide users to search using a query or keyword the systems thinking research papers from the repository. We leverage both semantic and vector based search to find and present the relevant research papers based on the paper metadata and the matches with the paper text embeddings. In addition to it, we also provide a shortcut for users to start exploring using the sustainable development goals. 

\subsection{Explorer}

\begin{figure}[ht]
    \centering
    \includegraphics[width=\textwidth]{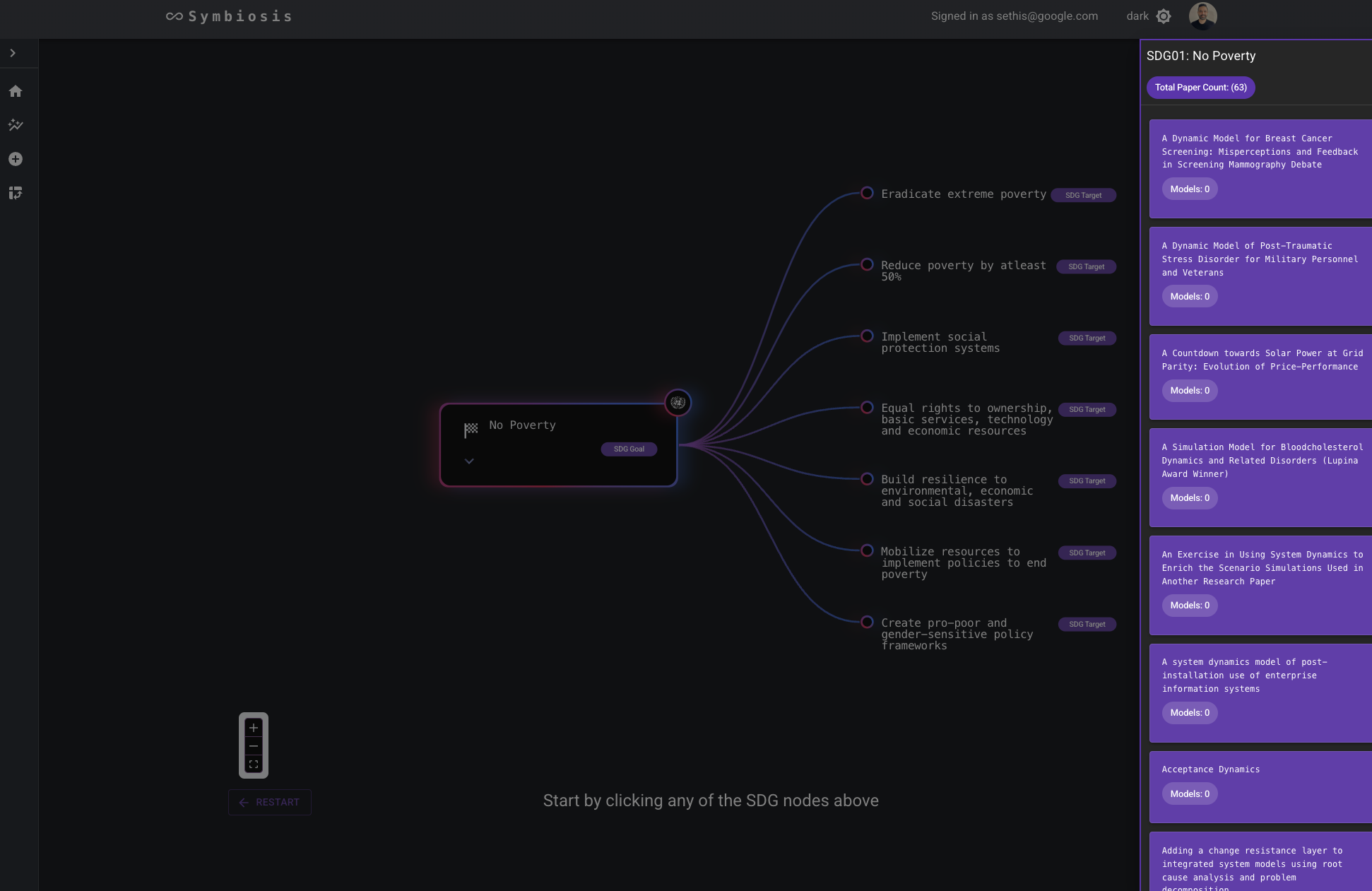}
    \caption{SYMBIOSIS Explorer Page}
    \label{fig:sdg_and_paper_explorer}
\end{figure}

The \textit{SYMBIOSIS} explorer provides a simple interface for exploring the contents of the repository via sustainable goals and indicators.  Selecting a SDG or indicator presents a side-drawer with the relevant papers in the repository and prioritize those papers that include a causal loop or stock and flow diagram (as shown in Figure \ref{fig:sdg_and_paper_explorer}).  Once a paper is selected, the associated model will be displayed and a generative AI co-pilot interface can be utilized to ask questions about the model and topic domain.

\subsection{Generative AI as a co-pilot}

\begin{figure}[ht]
    \centering
    \includegraphics[width=\textwidth]{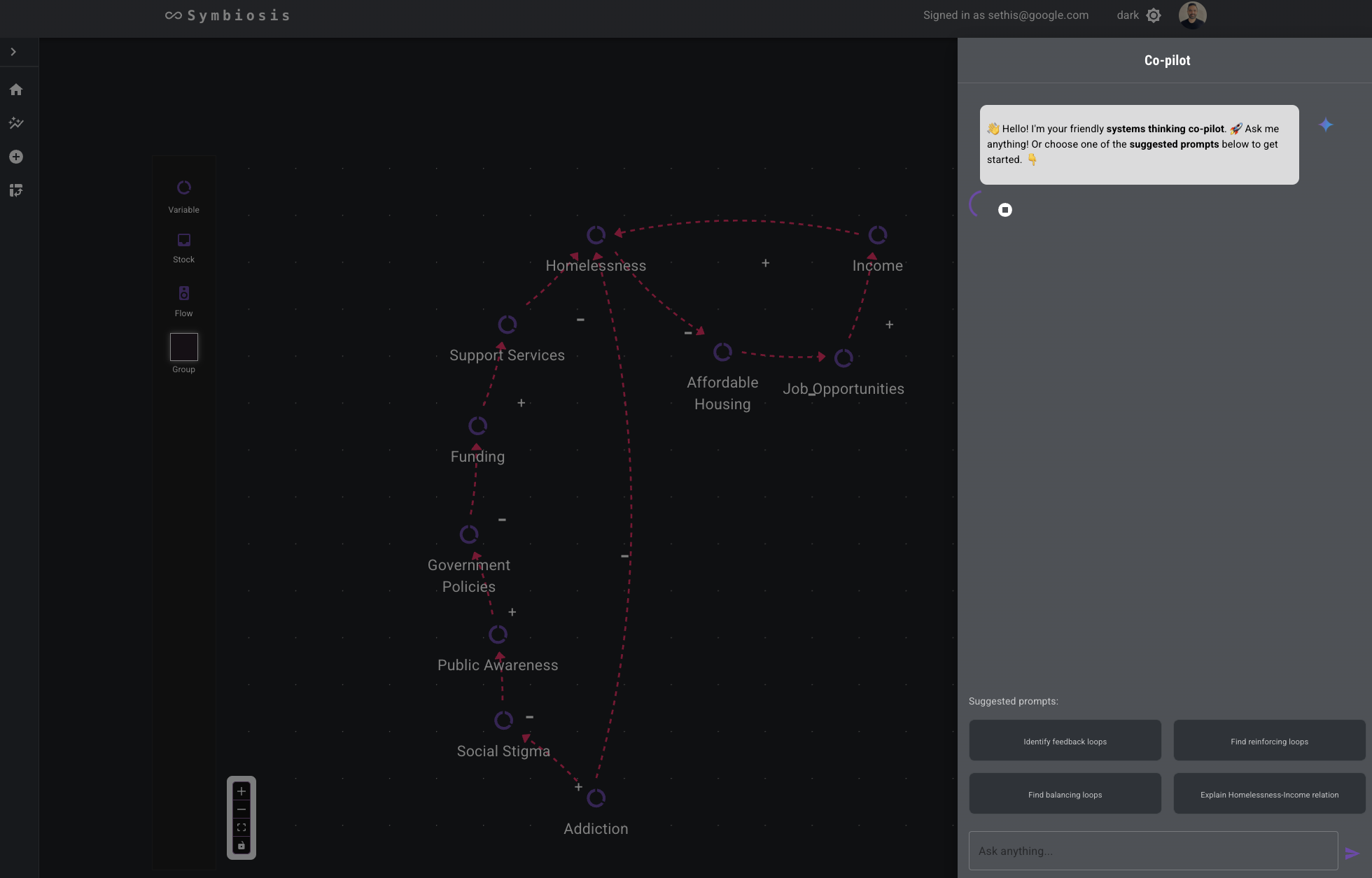}
    \caption{SYMBIOSIS Explorer Page}
    \label{fig:gemini_co_pilot}
\end{figure}

Finally, one of the key features of this platform is to enable users to explore existing research on systems thinking as well as create models from scratch without first learning the systems thinking/system dynamics notation, which often creates a big barrier for entry. Using natural language users can ask questions to the co-pilot to understand and explore the research papers and the models (as shown in Figure \ref{fig:gemini_co_pilot}) as well as leverage co-pilot to create and build upon models from scratch all the while describing the problem context in natural language. We have used Gemini \citep{geminiteam2024geminifamilyhighlycapable} as the LLM in this example, any other open-source or API-based private LLMs can be easily replaced as the LLM of choice.

\section{Conclusion}

This research presented \textit{SYMBIOSIS}, an AI-powered framework and platform designed to increase the accessibility and usability of Systems Thinking for addressing societal challenges. Our work offers a centralized repository of systems thinking models, categorized by SDGs, and a generative AI co-pilot to translate between diagrammatic and natural language representations. By bridging the gap between complex systems representations and user understanding, SYMBIOSIS aims to facilitate broader engagement with Systems Thinking and foster its integration into AI development for more responsible and effective societal problem-solving.

\subsection{Limitations}

This work identifies several limitations that offer avenues for future research and development in the AI for Systems Thinking space:

\begin{itemize}
    \item \textbf{XMILE Standard}: The current reliance on the XMILE standard for exchanging system dynamics models presents challenges due to the limitations of XML for handling complex mathematical formulas and the lack of robust validation mechanisms. We faced these problems while building the parser used to transformed XMILE models to structured data to be stored in a knowledge graph database for access via querying and vector search. This highlights the need for exploring alternative, more modern data exchange formats that are better suited for representing and validating the rich information contains in system dynamics models.
    \item \textbf{SDG Classifier}: The SDG classifier had limitations in precision and recall performance. This is likely due to the fact that the training data was generated using keyword-based queries, which can lead to noisy and incomplete labels. This suggests the need for development of a more comprehensive and accurately labeled training dataset based on expert knowledge and manual annotation.
    \item \textbf{Generative AI}: In leveraging generative AI to extract structured representations from images, we note that the space of multimodal models for diagrams is relatively limited, compared to creative image generation. This was empirically evident in the models performance. Also, multimodal outputs are not available in the version of Gemini Pro used, so we were not able to verify the model's claim that the object detection task was performed. Overall, smaller CNN models trained on this specific task would likely see improved performance until general models catch up.
    \item \textbf{Visualization Standardization}: The lack of standardized visualization metadata and layout algorithms for system dynamics models limits the accessibility and interpretability of these models for users. Since systems thinking and system dynamics tools do not require the inclusion of layout information, there is no consistent visual representation across different platforms. This inconsistency can make it difficult for users to understand and compare models. Future work should focus on developing standardized visualization guidelines and algorithms, potentially drawing inspiration from graph drawing techniques, to ensure consistent and user-friendly visual representations.
    \item \textbf{Evaluation of LLM co-pilot for Systems Thinking}: LLMs are very well known to hallucinate in pretty much all tasks whether it's related to performing data-to-text \citep{rebuffel2021controllinghallucinationswordlevel}, summarization \citep{maynez2020faithfulnessfactualityabstractivesummarization} or question-answering (QA) \citep{adlakha2024evaluatingcorrectnessfaithfulnessinstructionfollowing}. In this paper we have not performed any specific evaluation (other than cursory tests) to evaluate or benchmark the performance of co-pilot in the various tasks it is used for. We believe this as an important next step for the future work and form it's research on it's own.
\end{itemize}

\subsection{Future Work}

This work opens up several exciting possibilities for future research, such as:

\begin{itemize}
    \item \textbf{Investigating alternative data exchange formats}: Exploring modern data formats, such as RDF or JSON-LD, that offer better support for semantic representation, validation, and handling of mathematical expressions.
    \item \textbf{Improving the SDG classifier}: Expanding the training dataset with expert-validated examples, and incorporating contextual information beyond keyword-based matching.
    \item \textbf{Developing specialized tuned/foundation multimodal Generative AI models}: Creating and training models specifically designed for understanding and extracting information from system dynamic diagrams, or investing in developing a human annotated dataset to develop better accuracy CNN models to convert plethora of knowledge available in causal loop and stock-and-flow diagrams to structured knowledge graph will accelerate both AI for Systems Thinking and Systems Thinking for AI use-cases.
    \item \textbf{Establishing Visualization standard}: Defining clear guidelines and algorithms for the visual layout of systems thinking/system dynamics models to enhance their interpretability and facilitate comparitive analysis.
    \item \textbf{Expanding the scope of the co-pilot}: Enhancing the generative AI co-pilot to support a wider range of tasks, such as model validation, scenario analysis and generation, incrementally modifying an existing model using natural language and automated report generation.
\end{itemize}

Beyond improving the accessibility of Systems Thinking, this work lays the groundwork for enhancing AI/ML based products with systems thinking capabilities:

\begin{itemize}
    \item \textbf{Enhancing Problem Understanding and Formulation}: Integrate Systems Thinking principles into the AI development lifecycle, particularly during problem formulation and conceptualization. This can help AI models better understand complex relationships, feedback loops, and unintended consequences, leading to more robust and effective solutions. This is particularly crucial in high-stakes domains like healthcare, where understanding system-level effects is vital for safe and effective AI applications.
    \item \textbf{Benchmarking Generative AI for Systems Thinking}: Utilize the curated dataset of systems thinking models and the SDG classifier developed in this work to establish benchmarks for evaluating the causal and abductive reasoning abilities of generative AI models. This can involve tasks such as predicting system behaviour, identifying feedback loops, and generating explanations for observed phenomena.
    \item \textbf{Improving Conversational AI with Systems Thinking}: Leverage the knowledge graph extracted from system dynamics models to fine-tune generative AI models for improved conversational abilities on societal topics. This can involve training models to incorporate community perspectives, understand complex inter-dependencies, and generate responses that reflect a deeper understanding of systemic issues.
    \item \textbf{Developing "Systems of Thought" Prompting}: Expand on current research in Chain-of-Thought and other prompting techniques to explore novel "Systems of Thought" prompting methods. These techniques could guide generative AI models to reason about complex systems by explicitly considering feedback loops, time delays, and emergent behaviour. Evaluate the performance of these new prompting techniques against zero/few-shot and Chain-of-Thought prompting on societal understanding, explanation and reasoning tasks.
\end{itemize}

\bibliography{main}
\bibliographystyle{tmlr}

\end{document}